\documentclass[twocolumn,showpacs,preprintnumbers,amsmath,amssymb,aps,pre]{revtex4}
\usepackage{mathtools,amssymb,lipsum}

\usepackage{epsfig}

\usepackage{graphicx,latexsym,xcolor}

\usepackage{tikz}

\usepackage[english]{babel}
\usepackage{amsmath}
\usepackage{color}
\usepackage{epsfig}
\usepackage{graphicx}
\usepackage{bm}
\usepackage{times,amsmath,amssymb}
\usepackage{subfigure}
\usepackage{amsfonts}
\usepackage{amssymb}
\usepackage{color}
\global\long\def\ket#1{\left|#1\right\rangle }

\global\long\def\bege{\begin{equation}}
\global\long\def\ende{\end{equation}}

\global\long\def\begal{\begin{align}}
\global\long\def\endal{\end{align}}

\begin{document}

\title{From quantum to classical via crystallization}
\author{Ioannis Kleftogiannis, Ilias Amanatidis}
%\author{Ioannis Kleftogiannis$^1$, Ilias Amanatidis$^{2}$}
%\affiliation{$^1$ Physics Division, National Center for Theoretical Sciences, Hsinchu 30013, Taiwan }
%\affiliation{$^2$Department of Physics, Ben-Gurion University of the Negev, Beer-Sheva 84105, Israel}
%\affiliation{$^1$ Independent Researcher, email: ph04917@yahoo.com }
%\affiliation{$^2$ Independent Researcher, email: eliasamanatidis@hotmail.com}

\date{\today}
\begin{abstract}
We show that classical states can emerge as pure ground state
solutions of a quantum many-body system. We use a simple 
Hubbard model in 1D with strong short-range interactions
and a second nearest neighbor hopping with N particles arranged among M sites. We show that the ground state of this Hubbard chain for $M=2N-1$ consists of a single many-body state where the strongly interacting particles arrange in a classical state with crystalline order. The ground state is separated by an energy gap
from the first excited state, and survives
in the thermodynamic limit for large N. The energy gap
increases linearly with the strength of the interaction
between the particles making the classical ground state
robust to external perturbations like disorder. Our result is an example of how a quantum system can converge to a classical state, like a crystal, without requiring decoherence, wavefunction collapse or other external mechanisms.

\end{abstract}

%\pacs{03.65.Nk, 05.60.-k, 05.40.Fb, 72.15.Rn, 73.63.Nm}
\maketitle
The emergence of classical states of matter out
of quantum mechanical systems remains an important
problem in physics of fundamental significance \cite{namiki,zurek,tegmark,leslie,xu,schlosshauerall}.
Particularly elusive is the mechanism that gives rise 
to classical states of matter out of a collection
of particles behaving quantum mechanically, such as atoms, interacting via various forces.
One example is the formation
of a crystal via the process of crystallization\cite{uwaha,viedma,ben}
where atoms obeying quantum mechanics interact
via electric forces, such as Coulomb and Van Der Waals,
giving rise to a crystal, with the atoms arranged
in a lattice structure, that behaves
classically on the macroscopic level with robust universal
properties like rigidity. Particularly the perfection of the crystal  at larges scales is traditionally very hard to derive from first principles. In addition, usually the transition of a physical system from quantum to classical is investigated by taking the limit of the Planck's constant $\hbar \rightarrow 0$. Other external mechanisms like decoherence\cite{namiki,zurek,tegmark,leslie,xu,schlosshauerall} are also employed, where a random phase in the wavefunction results in destructive interference that eliminates the quantum effects. Moreover it is usually debated whether the wavefunction collapse via observation is related to the process of crystallization that turns a quantum system into a classical one.

Many-body effects in quantum systems can be studied
via various methods such as Hubbard models
\cite{cdw1,cdw2,cdw3,cdw4,masella,papers,slava}. 
For our study we consider N particles arranged among M sites in a Hubbard chain where only one particle is allowed per site, with Hamiltonian
\begin{equation}
H=U\sum_{i=1}^{M-1} n_{i}n_{i+1} + 
t\sum^{M-2}_{\substack{ \text{ i=1 } }}(c_{i}^{\dagger}c_{i+2} +h.c.),
\label{eq_h_1d}
\end{equation}
where $c^{\dagger}_{i},c_{i}$ are the creation 
and annihilation operators for a particle at site i in the chain and $n_{i}=c_{i}^{\dagger}c_{i}$ is the number operator,
taking the value $0(1)$ for an empty(occupied) site.
For our calculations we fix the value of the hopping at $t=1$.
The model Eq. \ref{eq_h_1d} could describe for example hard-core bosons
satisfying the commutation relation $[c_{i},c_{j}^{\dagger}] = (1-2n_{i})\delta_{ij}$.

For large interaction strengths $U \gg t$ and odd total number of sites M in the chain, the ground state of the model Eq. \ref{eq_h_1d} is a quantum fluid (superfluid) phase where the particles arrange in clustering structures that minimize the energy of the system, characterized by a fixed number of clusters\cite{papers}. For $M>2N \Rightarrow f<\frac{1}{2}$ the clusters are single particles with zero cluster length and the resulting superfluid phases consist of charge-density-wave(CDW) states, where the particles do not occupy adjacent sites in the Hubbard chain. Restricting the empty space in the system, for $M=2N-1$, gives one ground state with crystalline order of the type $\ket{1010101...01}$ at energy E=0, as if the particles were frozen on odd sites of the Hubbard chain. For larger fillings $M<2N-1 \Rightarrow f>\frac{N}{2N-1}$ the particles condense into  a fixed number of clusters with variable lengths.

The crystalline state can be demonstrated 
by considering a small system described by Eq. \ref{equation1} with N=2,M=3 containing three Fock states $\ket{101},\ket{110},\ket{011}$.
The corresponding Hamiltonian matrix of the system is
\begin{equation}
H=\left( \begin{array}{ccc}
0 & 0 & 0  \\
0 & U & t  \\
0 & t & U  \\
\end{array} \right).
\label{toy1}
\end{equation}
The eigenvalues of the above Hamiltonian are $E_0=0, E_1=U+t, E_2=U-t$. For $U>t$ the ground state of the system is the single Fock state
 $\ket{101}$ with energy $E_0=0$, separated by an energy gap $U-t$
 from the first excited state.
%%%%%%%%%%%%%%%%%%%%%%%%%5
%%%%%%%%%%%%%%%%%%%%%%%%%

\begin{figure}
\begin{center}
\includegraphics[width=0.9\columnwidth,clip=true]{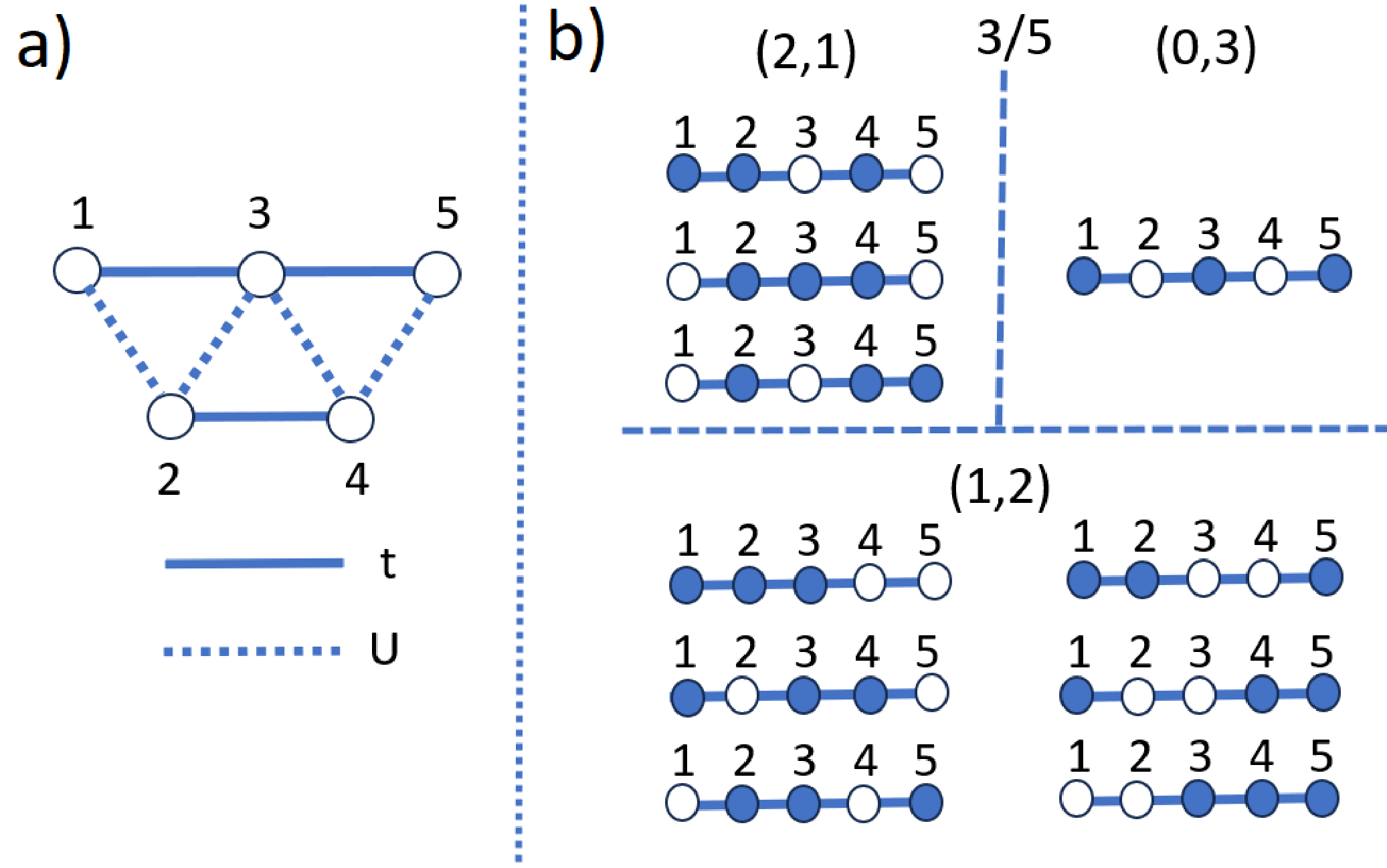}
\end{center}
\caption{a) The schematic of the Hubbard chain that we use. b) An example of the possible Fock states, forming the Hibert space of the system, with N=3 particles distributed among M=5 sites. Because of the second-nearest-neighbor hopping, the Hilbert space is split into three sectors according to the number of even and odd occupied sites in the chain, denoted by $(even,odd)$. The corresponding Hamiltonian of the chain can be written in a block-diagonal form, since only Fock states in the same sector can communicate via the second-nearest-neighbor hopping. Each sector of the Hilbert space corresponds to a Hamiltonian block whose size is determined by the number of Fock states in the sector.}
\label{fig1}
\end{figure}

The hopping term t in Eq. \ref{eq_h_1d} allows the particles to hop only between second-nearest-neighboring sites in the Hubbard chain. By enumerating the sites as even and odd, we can see that the particles will stay on the same type of sites during the hopping process. Therefore Fock states with a different number of particles on even and odd sites cannot communicate via the hopping. This allows to split(decompose) the Hamitlonian Eq. \ref{eq_h_1d} in blocks, resulting in block-diagonal form that allows us to diagonalize it more efficiently for a large number of particles N. Each block corresponds to Fock states that have the same number of particles on even and odd sites which can be denoted as $(even, odd)$ with $even+odd=N$. The total number of blocks can be calculated by finding the number of compositions of N into two parts. Since the blocks are given by $(N-k, k)$ for $k=1,2,...,N$ we get N blocks. Notice that for $M=2N-1$ the case $(N,0)$ is excluded, since it contains particles only on even sites. In addition the size of each block is given by
\begin{equation}
D_{k}(M,N) =  \binom{\frac{M-1}{2}
}{N-k}\binom{\frac{M+1}{2}
}{k}.
\end{equation}
Summing the above equation over N we get
the total number of Fock states
\begin{equation}
\sum_{k=1}^{N}D_{k}(M,N) =  \binom{M
}{N}.
\end{equation}
We show an example of the block decomposition method in Fig. \ref{fig1} for a small system with N=3,M=5. The Hilbert space of the system can be split into the blocks $(2,1),(1,2),(0,3)$. The block $(0,3)$ contains just one Fock state, the classical crystal state, where the particles occupy odd sites and are separated by one empty site. The rest of the blocks contain Fock states with clustering in the particle structures. 
%%%%%%%%%%%%%%%%%%%%%%%%%%%%%%%%5
%%%%%%%%%%%%%%%%%%%%%%%%%%%%%%%%%%%

In general, for $M=2N-1$ there is always one block $(0,N)$ consisting of a single zero matrix element, corresponding to the eigenvalue(energy) $E=0$ with eigenstate $\ket{1010101...01}$, a classical state with crystalline order of periodicity $2\alpha$ where $\alpha$ is the lattice constant. When there are no negative eigenvalues this state becomes the ground state of the system. We investigate when this is realized for different number of particles $N$ and values of the interaction strength $U$.

We have found that the first excited state corresponds to the lowest eigenvalue of the Hamiltonian block $(1,N-1)$, which we denote by $E_1$. In addition for $U=0$, the Hamiltonian matrix of this block can be mapped into the Hamiltonian matrix of a single particle hopping on a tight-binding square lattice of size $M_x \times M_y$ with $M_x=\frac{M+1}{2}=N$ and $M_y=\frac{M-1}{2}=M-N$ with first-nearest-neighbor hopping $t=1$. The sites of the square lattice correspond essentially to the Fock states and the hopping between the sites corresponds to the second-nearest-neighbor hopping term in Eq \ref{eq_h_1d}. Therefore the energy spectrum of the block $(1,N-1)$ is given by the energy spectrum of the tight-binding lattice
\begin{equation}
E(j_x,j_y) = 2cos\left(\frac{\pi j_x}{M_x+1}\right)+2cos\left(\frac{\pi j_y}{M_y+1}\right),
\end{equation}
where $j_x=1,2,...,M_x$ and  $j_y=1,2,...,M_y$. 
Switching on the interaction between the particles U 
introduces an on-site potential in the square lattice model
that follows the pattern
\begin{equation}
\begin{aligned}
& \{ U,U \} && \text{for } M=3 \\
& \{ 2U,U,U,U,U,2U \} && \text{for } M=5 \\
& \{ 2U,2U,U,2U,U,U,U,U,2U,U,2U,2U \} && \text{for } M=7.
\end{aligned}
\end{equation}
The above pattern can be expressed as a recurrence relation in the following way
\begin{equation}
\begin{split}
C_M&=\bigl\{ \bigl\{ \underbrace{2U,..,2U}_{\text{$\frac{M-3}{2}$ }},U \bigl\} \bigl\}  \cup C_{M-2}\cup \bigl\{{U,\underbrace{2U,..,2U}_{ \text{$\frac{M-3}{2}$ }}} \bigl\}: \\  &   M\geq3 \bigl\}
\end{split}
\end{equation}
for $C_1=\bigl\{\varnothing \bigl\}$ and $M$ odd. 
%%%%%%%%%%%%%%%%%%%%%%%%%%%%%%%%%%%%%%%%%%%%%%%%%%%%%%%%%%%%%

\begin{figure}
\begin{center}
\includegraphics[width=0.9\columnwidth,clip=true]{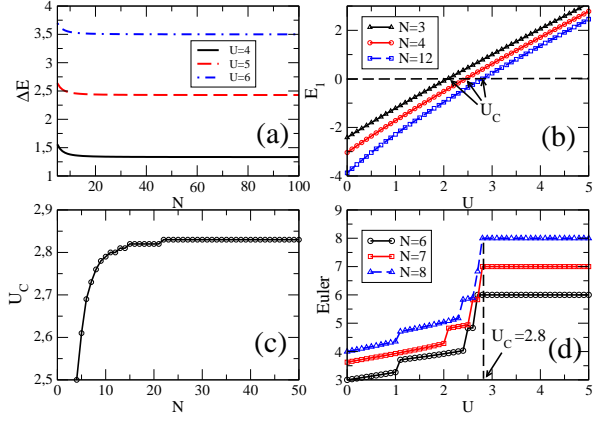}
\end{center}
\caption{a) The energy gap $\Delta E$ from the ground state to the first excited state of the Hubbard chain for different interaction strengths U between the particles. The gap converges to finite values for a large number of particles N, at the thermodynamic limit. b)The value of the first excited state $E_1$ increases linearly with U for $U \gg t$, for all N. The value of U where $E_1=0$, determines the critical value $U_C$, where the system transitions from a quantum supefluid phase to a classical crystalline phase. c) The critical interaction strength $U_C$ converges for large N. d) The Euler characteristic, describing the clustering of the interacting particles, forms of a plateau at $\chi=N$ for $U>U_C$, when the system transitions to the classical crystal.}
\label{fig2}
\end{figure}

We calculate the energy gap $\Delta E = E_1-E_0$ from the ground state $E_0=0$ to the first excited state $E_1$. As can be seen in Fig. \ref{fig2}a the gap  converges to finite values for a large number of particles N, at the thermodynamic limit, for different values of the interaction strength U. In Fig. \ref{fig2}b we show how the lowest eigenvalue $E_1$ of the $(1,N-1)$ block changes with U, for the three curves corresponding to different N. The point where the curves cross the $y=0$ axis, is the critical interaction strength $U_{C}$ where a phase transition occurs and the crystalline state at energy $E=0$ becomes the ground state of the system, for $U>U_C$. For $U<U_C$ the negative eigenvalue $E_1$ is the ground state energy, corresponding to different superfluid phases. In addition we notice that $E_1$ and consequently the gap $\Delta E$ increases linearly with U as $\Delta E=\alpha U+\beta$ for $U>U_C$, where $U_{C}=\frac{-\alpha}{\beta}$. In Fig. \ref{fig2}c we show that the critical value $U_C$ converges for large $N$. In Fig. \ref{fig2}d we plot the Euler characteristic of the structures formed by the particles, defined as $\chi=\sum \Psi_i^2 \chi_i$ where $\Psi_i^2$ is the probability of each Fock state i in the ground state and $\chi_i=N-L_i$ is the corresponding Euler characteristic, where $L_i$ is the number of bonds between the particles, formed when two of them occupy adjacent sites in the Hubbard chain. A quantum phase transition is shown at $U_C$, where the Euler characteristic forms a plateau at $\chi=N$ for $U>U_C$, as the system transitions from a superfluid phase for $U<U_C$, to the classical crystal state which has no bonds between the particles $(L_i=0)$ for $U>U_C$. This crystal state has broken translational symmetry, which is satisfied by the superfluid phases appearing for $U<U_C$. 

These additional phases correspond to the non-flat steps of the Euler characteristic for $U<U_C$. We have found that for $M \ge 2N-1$ the number of these phases/steps is associated to the number of blocks in the block-diagonal Hamiltonian, via the integer partitions of N into two parts as
\begin{equation}
S(N) =  1 + \frac{1}{4} \left( 2N -1 + (-1)^{N} \right).
\end{equation}
Each of these phases is related to a symmetry in the Fock states
belonging to specific blocks. For example we can notice
that for every Fock state inside the blocks (1,2) and (2,1) in Fig. \ref{fig1}b its mirror symmetric state appears also.
%If we denote each Fock state by the occupied sites in the Hubbard %chain as $\ket{x(1),x(2),..,x(N)}$ where $x(i)$ is the position of %the ith particle, the permutations $x(i) \rightarrow M-x(i)-1$ for %$x(i) < \frac{M-1}{2}$ and $x(i) \rightarrow M-x(i)+1$ for $x(i) > %\frac{M-1}{2}$ give all the states in the block.
Applying the permutation $i \leftrightarrow M-i+1$ for $i \le \frac{M-1}{2}$ for the state (occupied(1) or empty(0)) of the ith site in the Hubbard chain, on each Fock state in a block gives its mirror symmetric state. Therefore the superfluid phase corresponding to the blocks (1,2) and (2,1) in Fig. \ref{fig1}b, contains a parity symmetry. Breaking such symmetries in the Hilbert space of the system results in different superfluid phases, corresponding to the different non-flat steps in the Euler characteristic for $U<U_C$ in Fig. \ref{fig1}d.
%%%%%%%%%%%%%%%%%%%%%%%%%%%%%%%%%%
%%%%%%%%%%%%%%%%%%%%%%%%%%%%%%%%%%%

As another simple example of the crystallization mechanism
analyzed above, consider a Hubbard chain with two bosons arranged among two sites with Hamiltonian
\begin{equation}
H=U n_{1}n_{2} + 
t (c_{1}^{\dagger}c_{2} + c_{2}^{\dagger}c_{1}).
\label{eq_h_boson}
\end{equation}
When both bosons occupy the same site they interact
with energy U, lifting the energy of the system. 
The bosons can also hop between the two sites via a hopping t.
There are three possible configurations of the particles
among the sites, the Fock states $\ket{20},\ket{02},\ket{11}$. The matrix form of the Hamiltonian in the basis of these Fock states is
\begin{equation}
H=\left( \begin{array}{ccc}
U & 0 & t  \\
0 & U & t  \\
t & t & 0  \\
\end{array} \right).
\label{equation1}
\end{equation}
The eigenvalues of the above Hamiltonian are $E_{0}=U, E_{1}=\frac{1}{2}(U-\sqrt{8t^2+U^2}, E_{2}=\frac{1}{2}(U+\sqrt{8t^2+U^2})$. At the limit of strong interaction between the particles $U \gg t$,
we have $E_{0}=U, E_{1}=0, E_{2}=U$, where the energy $E_{1}=0$ corresponds to the eigenstate $\ket{11}$. This is the ground state of the system separated by an energy gap U from the two other degenerate states with energy $E_{0}=E_{2}=U$. This is another simple example of a classical ground state with crystalline order where the crystal periodicity is simply the lattice constant.
%emerging as a pure ground state solution of a quantum many-body system, for strong interaction strength between the particles. 

%%%%%%%%%%%%%%%%%%%%%%%%%%%%%%%%
\begin{figure}
\begin{center}
\includegraphics[width=0.9\columnwidth,clip=true]{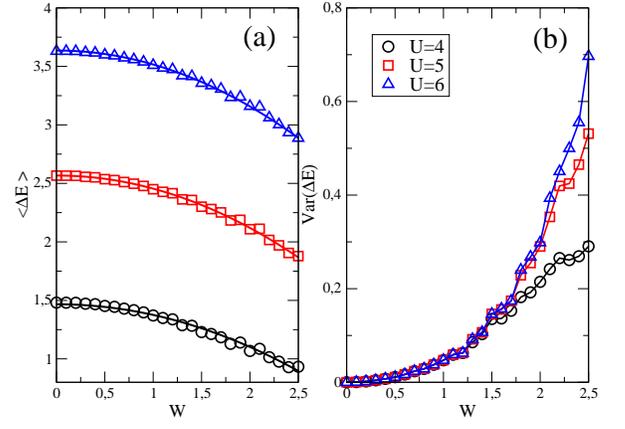}
\end{center}
\caption{a) The mean value of the energy gap $\langle \Delta E \rangle$ versus the disorder strength W for N=6 number of particles and different interaction strengths U. All the points can be fitted with fitting curves(solid lines) $a+bW^2$, with a,b being two fitting parameters. b) The corresponding variance of the energy gap.}
\label{fig3}
\end{figure}
%%%%%%%%%%%%%%%%%%%%%%%%%%%%%%%%%%%%%%%%%

To investigate the robustness of the energy gap
we introduce disorder in our model, by considering an on-site random potential $H_d=\sum_{i=1}^{M} V_i n_{i}$ in the Hamiltonian Eq. \ref{eq_h_1d}. The Hamiltonian of a small system with N=2,M=3 is 
\begin{equation}
H=\left( \begin{array}{ccc}
V_1+V_3 & 0 & 0  \\
0 & U+V_1+V_2 & t  \\
0 & t & U+V_2+V_3  \\
\end{array} \right).
\label{equation1}
\end{equation}
The eigenvalues of the above matrix are
\begin{equation}
 \begin{split}
        E_{0} &= V_{1}+V_{3} \\
        E_{1}&=\frac{1}{2} \left( 2U+V_{1}+2V_{2}+V_{3}-\sqrt{4 t^{2}+(V_{1}-V_{3})^{2} }  \right)\\
        E_{2} &=\frac{1}{2}
 \left(2U+V_{1}+2V_{2}+V_{3}+\sqrt{4 t^{2}+(V_{1}-V_{3})^{2} }  \right)
    \end{split}.
\label{disorder01}    
\end{equation}
The energy gap $ \Delta E = E_1 -E_0 $ from the ground state to the first excited state is
\begin{equation}
 \begin{split}
       \Delta E&= \frac{1}{2} \left( 2U-V_{1}+2V_{2}-V_{3}-\sqrt{4 t^{2}+(V_{1}-V_{3})^{2} }  \right) \\
\end{split}.
\label{disorder02}    
\end{equation}
In order to calculate the mean value of the energy gap $ \langle \Delta E \rangle = \langle E_1 -E_0 \rangle $ we consider that the random numbers $V_1,V_2,V_3$ follow the box probability distribution $P(V)=\frac{1}{W}$, in the range $[-\frac{W}{2},\frac{W}{2}]$ with $W$ being the strength of the disorder, with mean value $\langle V \rangle =0$. Since $\langle V_1 \rangle = \langle V_2 \rangle = \langle V_3 \rangle =0 $ the mean value of Eq. \ref{disorder02} is
\begin{equation}
 \begin{split}
       \langle \Delta E \rangle &=\left( U-\frac{\langle \sqrt{4 t^{2}+(V_{1}-V_{3})^{2}} \rangle}{2}  \right) \\
\end{split}.
\label{disorder03}    
\end{equation}
The square-root term can be written as
\begin{equation}
 \begin{split}
        \sqrt{4 t^{2}+(V_{1}-V_{3})^{2} } &=  2t\sqrt{1+\frac{(V_{1}-V_{3})^{2}}{4 t^{2}} } \\
    \end{split}.
\label{disorder04}    
\end{equation}
We can apply a Taylor expansion $(1+x)^{a}=1+ax$ in the above expression with $x=\frac{(V_{1}-V_{3})^{2}}{4 t^{2}}$ and $a=\frac{1}{2}$ by assuming weak disorder with $W \ll 1$ and therefore $|x|<1$ and $|ax|\ll1$. We have 
\begin{equation}
 \begin{split}
        2t\sqrt{1+\frac{(V_{1}-V_{3})^{2}}{4 t^{2}} } =  2t\left(1+\frac{(V_{1}-V_{3})^{2}}{8 t^{2}} \right)
    \end{split}.
\label{disorder1}    
\end{equation}
The mean value of the term containing the random numbers is 
\begin{equation}
 \begin{split}
       \langle (V_1- V_3)^2 \rangle =\langle V_1^2 \rangle + \langle V_3^2 \rangle -2 \langle V_1V_3 \rangle
    \end{split}.
\label{disorder2}    
\end{equation}
Since the random numbers follow the box distribution
$P(V)=\frac{1}{W}$ the mean values can be calculated
via the integrals
\begin{equation}
 \begin{split}
        \langle V^2 \rangle & = \int_{-\frac{W}{2}}^{\frac{W}{2}} V^2 \frac{1}{W} \,dV =\frac{W^2}{12}  \\
        \langle VV^{'} \rangle & = \int_{-\frac{W}{2}}^{\frac{W}{2}} \int_{-\frac{W}{2}}^{\frac{W}{2}} VV^{'} \frac{1}{W} \,dVdV^{'}  = 0
    \end{split}. 
    \label{disorder4} 
\end{equation}
After plugging the above results into Eq. \ref{disorder2} we obtain,
\begin{equation}
 \begin{split}
       \langle (V_1- V_3)^2 \rangle = \frac{W^2}{6}
    \end{split}. 
    \label{disorder5} 
\end{equation}
By taking account of Eq. \ref{disorder1} then Eq. \ref{disorder03} becomes
\begin{equation}
 \begin{split}
        \langle \Delta E \rangle= U - t -\frac{W^2}{48t}
    \end{split}.
\label{disorder6}    
\end{equation}
At the limit of weak disorder $W \ll t$ and
strong interaction strength $U \gg t$, the 
above equation gives $\langle \Delta E \rangle= U$,
that is, the energy gap is not affected by the
disorder. We expect that the energy gap
remains robust to disorder for a larger
number of particles also.

In Fig. \ref{fig3}a we show the mean gap $\langle \Delta E \rangle$
vs the disorder strength W for N=5 and different interaction strengths U. All the points can be fitted with fitting curves 
of the type $a+bW^2$, with a,b being two fitting parameters.
Therefore the form of Eq. \ref{disorder6} is generic for all N.
In addition the variance of the gap in Fig. \ref{fig3}b
is negligible for weak disorder($W<0.5$). Therefore, the energy gap
from the classical crystalline state to the first excited state is robust to external perturbations like disorder.

To conclude we have shown that classical states
can emerge as ground state solutions of a quantum many-body system.
We have demonstrated the effect for a Hubbard chain with N particles
distributed among M=2N-1 sites, with a second-nearest-neighbor hopping
t and strong nearest-neighbor interaction U. The ground state of this system consists of one classical state where the particles
do not occupy adjacent sites in the Hubbard chain, arranging in a crystal, minimizing the energy of the system. The energy gap from
the first excited state remains finite at the thermodynamic limit
for large N. In addition it increases linearly with U,
making the system robust to external perturbations like disorder,
as we have demonstrated via numerical and analytical calculations.
Our result shows that classical states can emerge as pure states
out of a quantum mechanical system, without employing external mechanics like decoherence or wavefunction collapse.
The crystallization mechanism could be particularly relevant to the formation of 1D molecular structures such as proteins.
Apart from its fundamental significance our result could
be also realized in cold-atoms experiments.

\section*{References}

%%%%%%%%%%%%%%%%%%%% Bell

\end{document}